\documentclass[aps,prc,twocolumn,superscriptaddress,floatfix,nofootinbib]{revtex4-2}

\usepackage{amsmath,amssymb,bm}
\usepackage{graphicx}
\usepackage{booktabs}
\usepackage{hyperref}
\hypersetup{hidelinks}
\usepackage{microtype}

\newcommand{\cs}{\mathrm{CSM}}
\newcommand{\td}{\mathrm{TD}}
\newcommand{\init}{\mathrm{init}}
\newcommand{\finalh}{\mathrm{final}}
\newcommand{\surv}{\mathrm{surv}}

\begin{document}

\title{Time-evolution formalism in the complex scaling method: Application to the two-proton decay of $^{6}$Be}

\author{Yuma Kikuchi}
\email{kikuchi@tokuyama.ac.jp}
\affiliation{Tokuyama College, National Institute of Technology, Shunan 745-8585, Japan}
\affiliation{RIKEN Nishina Center, 2-1 Hirosawa, Wako 351-0198, Japan}

\author{Kiyoshi Kat\={o}}
\affiliation{Nuclear Reaction Data Centre, Faculty of Science, Hokkaido University, Sapporo 060-0810, Japan}

\author{Takayuki Myo}
\affiliation{General Education, Faculty of Engineering, Osaka Institute of Technology, Osaka 535-8585, Japan}
\affiliation{Research Center for Nuclear Physics (RCNP), Osaka University, Ibaraki 567-0047, Japan}

\date{\today}

\begin{abstract}
\begin{description}
\item[Background]
Two-proton decay provides a direct probe of continuum dynamics and proton-proton correlations in proton-rich nuclei. A microscopic description that follows the decay in real time while retaining the relevant three-body correlations is essential for clarifying how the spatial and spin correlations of the emitted protons evolve during the decay.

\item[Purpose]
We apply our complex-scaled time-evolution operator to the two-proton decay of $^{6}$Be. Using an explicit $\alpha+p+p$ three-body description with Jacobi-coordinate rearrangement, we examine the relation between the decay width obtained from explicit time evolution and the CSM resonance-pole width and investigate the time-dependent spatial geometry and spin correlations of the emitted protons.

\item[Method]
The $^{6}$Be nucleus is described as an $\alpha+p+p$ three-body system with explicit Jacobi-coordinate rearrangement. The proton-proton subsystem is treated with the realistic Argonne $v8'$ NN interaction, which contains strong short-range repulsion. A confined $0^+$ initial state is expanded over the complex-scaled eigenstates of the final Hamiltonian and propagated with our complex-scaled time-evolution operator. The decay width is extracted from the survival probability. The projected decay component is analyzed in the $T$- and $Y$-type Jacobi coordinates using two-dimensional densities, mass-scaled hyperradial and hyperangular distributions, and spin decomposition.

\item[Results]
The late-time survival probability gives $\Gamma_{\rm TD}=0.107$ MeV, in close agreement with the CSM pole width $\Gamma_{\rm CSM}=0.106$ MeV, whereas the short-time nonexponential behavior is sensitive to how the initial confined state is prepared. The norm loss reflects the continuous removal of outgoing strength in the complex-scaled representation and depends only weakly on the scaling angle over the range considered. In the $T$-type coordinates, the small-$\beta$ component remains about $80\%$--$84\%$ from $t=100$ to $1000$ fm/$c$, indicating that a correlated two-proton configuration persists during the outward propagation. In the $Y$-type coordinates, the small-$\beta$ component associated with a relatively compact $\alpha+p$ geometry decreases from about $8\%$ to about $4\%$, while the large-$\beta$ three-body component becomes increasingly dominant. The projected decay component is strongly dominated by the spin-singlet sector, and the CHSH indicator approaches $2.61$ at late times.

\item[Conclusions]
The complex-scaled time evolution consistently reproduces the resonance width while resolving the development of the three-body decay geometry in real space. The explicit Jacobi-coordinate rearrangement reveals complementary aspects of the decay geometry: the correlated two-proton component remains dominant during the decay, whereas the relatively compact $\alpha+p$ configuration becomes progressively less important. The persistence of the correlated two-proton configuration, despite the strong short-range repulsion of the realistic Argonne $v8'$ NN interaction, is consistent with the qualitative picture obtained in previous time-dependent studies and extends this picture to the present three-body description.
\end{description}
\end{abstract}

\maketitle

\section{Introduction}

Unbound nuclei provide an important testing ground for understanding the interplay between intrinsic nuclear structure and continuum dynamics. In proton-rich systems, two-proton ($2p$) emission is particularly interesting because its dynamics reflect the proton-proton nuclear interaction, Coulomb repulsion, and correlations already present in the parent state. Consequently, $2p$-decay observables can provide information not only on the resonance properties of the parent nucleus but also on the spatial and spin correlations between the emitted protons \cite{Blank2008,Pfutzner2012,Pfutzner2023}.

$^{6}$Be is a simple and well-studied system for investigating correlated $2p$ decay microscopically. A natural description is provided by an $\alpha+p+p$ three-body model, in which the $0_1^+$ ground state lies in the three-body continuum. Since its $2p$-decay energy lies below the centroid of the $^{5}$Li resonance, $^{6}$Be is commonly regarded as a true two-proton emitter, for which sequential one-proton emission is energetically suppressed. However, because the $^{5}$Li ground-state resonance is broad, contributions associated with relatively compact $\alpha+p$ configurations are not completely excluded. The decay has therefore also been discussed in terms of democratic three-body decay, in which no well-separated sequential path dominates the final-state correlations \cite{Grigorenko2009,Grigorenko2012,Egorova2012,Chudoba2018,Oishi2014}. These features make $^{6}$Be a benchmark system for investigating the interplay between two-proton correlations and three-body decay dynamics.

Previous time-dependent studies by Oishi and collaborators have provided detailed analyses of proton-proton correlations in the $2p$ decay of $^{6}$Be \cite{Oishi2014,Oishi2017,Oishi2025}. In these calculations, the valence protons are described in a core-centered single-particle representation, and the proton-proton interaction is treated with an effective pairing interaction. More generally, Wang and Nazarewicz developed a time-dependent approach to two-nucleon emission and demonstrated the interplay between initial-state nucleon-nucleon correlations and final-state interactions \cite{Wang2021}. These studies established the importance of dynamical nucleon-nucleon correlations and showed that the predicted decay properties can depend on both the initial-state structure and the treatment of the final-state interaction.

A complementary perspective has been provided by the Gamow coupled-channel approach of Wang and collaborators, formulated in Jacobi coordinates \cite{Wang2017}. This work highlighted the role of genuine three-body configurations beyond a simple core-centered single-particle representation and showed that different Jacobi arrangements provide a natural framework for describing correlations involving the valence nucleons and the core.

Together, these studies motivate a more explicit three-body test of whether the correlated two-proton picture obtained in core-centered single-particle descriptions persists when the relevant Jacobi rearrangement channels are treated explicitly and the proton-proton subsystem is described with a realistic NN interaction containing strong short-range repulsion.

The complex scaling method (CSM) provides a unified description of bound, resonant, and nonresonant continuum states in an $L^2$ basis representation \cite{Aguilar1971,Balslev1971,Aoyama2006,Myo2014,MyoKato2020}. In our preceding work, we formulated a complex-scaled time-evolution operator based on the extended completeness relation (ECR) and applied it to continuum dynamics following the electric dipole excitation of $^{6}$He \cite{KikuchiTimeEvolution}. In this formulation, the time evolution is expressed through the spectrum of complex-scaled eigenstates, with resonant and nonresonant continuum components included on the same footing. The present study extends this framework to an explicit three-body decay process.

In the present work, we investigate the $2p$ decay of $^{6}$Be within an $\alpha+p+p$ three-body description that explicitly includes the rearranged Jacobi-coordinate channels and employs the realistic Argonne $v8'$ NN interaction for the proton-proton subsystem \cite{Pudliner1997}. The strong short-range repulsion of this interaction provides a stringent test of whether the correlated two-proton picture found in previous time-dependent calculations remains valid. We combine this three-body description with our complex-scaled time-evolution operator to follow the decay dynamics and analyze the spatial and spin correlations of the emitted protons. Thus, the central feature of the present calculation is the simultaneous use of explicit Jacobi rearrangement, the realistic Argonne $v8'$ NN interaction, and complex-scaled time evolution. The remaining interactions in the three-body Hamiltonian are specified in Sec.~III.

The initial state is prepared as a weakly bound $0^+$ state by modifying only the phenomenological three-body interaction. At $t=0$, the Hamiltonian is switched to the final Hamiltonian that reproduces the physical $0_1^+$ resonance of $^{6}$Be. The initial wave function is then expanded over the complex-scaled eigenstates of the final Hamiltonian and propagated with the complex-scaled time-evolution operator. The survival probability is used to extract the decay width, which is compared directly with the width obtained from the CSM resonance pole. The emitted component is further analyzed using the different rearranged Jacobi-coordinate representations and decomposed into spin-singlet and spin-triplet components.

The paper is organized as follows. Section~II presents the formulation of the complex-scaled time-evolution operator. Section~III describes the $\alpha+p+p$ three-body model and the preparation of the initial and final Hamiltonians. Section~IV defines the survival probability, projected decay density, and spin observables. Section~V presents the numerical results, including the decay-width comparison, spatial decay dynamics, and spin correlations. Section~VI summarizes the present work.

\section{Time-evolution operator in the complex scaling method}

In the CSM, the relative coordinates are transformed as
\begin{equation}
 \bm{r}\rightarrow \bm{r}e^{i\theta},
\end{equation}
where $\theta$ denotes the scaling angle. The transformation is implemented by the complex-scaling operator $\hat U(\theta)$, defined by $\hat U(\theta)\bm r\hat U^{-1}(\theta)=\bm r e^{i\theta}$ for each relative coordinate $\bm r$. The complex-scaled Hamiltonian is then defined by
\begin{equation}
 \hat H^\theta=\hat U(\theta)\hat H\hat U^{-1}(\theta),
\end{equation}
and its eigenvalue problem is
\begin{equation}
 \hat H^\theta |\Psi_\nu^\theta\rangle
 =E_\nu^\theta |\Psi_\nu^\theta\rangle.
\end{equation}
The states biorthogonal to $|\Psi_\nu^\theta\rangle$ are denoted by $\langle\widetilde{\Psi}_\nu^\theta|$. In the finite $L^2$ model space used in the present calculation, the complex-scaled eigenstates satisfy the extended completeness relation (ECR)
\begin{equation}
 \sum_\nu |\Psi_\nu^\theta\rangle
 \langle\widetilde{\Psi}_\nu^\theta|=1^\theta,
 \label{eq:ecr}
\end{equation}
where $1^\theta$ denotes the identity operator in the complex-scaled model space. The sum includes the bound states, isolated resonances, and discretized nonresonant continuum states retained in the model space.

The time-dependent wave function is formally written as
\begin{equation}
 |\Phi(t)\rangle=e^{-i\hat Ht/\hbar}|\Phi(0)\rangle.
\end{equation}
Here, $|\Phi(0)\rangle$ denotes the normalized confined initial state at $t=0$.
Using the CSM and Eq.~(\ref{eq:ecr}), the time-dependent wave function can be represented as
\begin{equation}
 |\Phi(t)\rangle=\hat U^{-1}(\theta)\sum_\nu |\Psi_\nu^\theta\rangle e^{-iE_\nu^\theta t/\hbar}\langle\widetilde{\Psi}_\nu^\theta|\hat U(\theta)|\Phi(0)\rangle.
 \label{eq:timeevolution}
\end{equation}
We define the expansion amplitude as
\begin{equation}
 C_\nu(0)=\langle\widetilde{\Psi}_\nu^\theta|\hat U(\theta)|\Phi(0)\rangle,
 \label{eq:initial_amplitude}
\end{equation}
so that the time evolution is expressed as a superposition of the complex-scaled eigenstates, each propagated with the phase factor $e^{-iE_\nu^\theta t/\hbar}$.

\section{Three-body model of $^{6}$Be}

\subsection{Three-body model and Gaussian expansion}

\begin{figure}[t]
\centering
\includegraphics[width=\columnwidth]{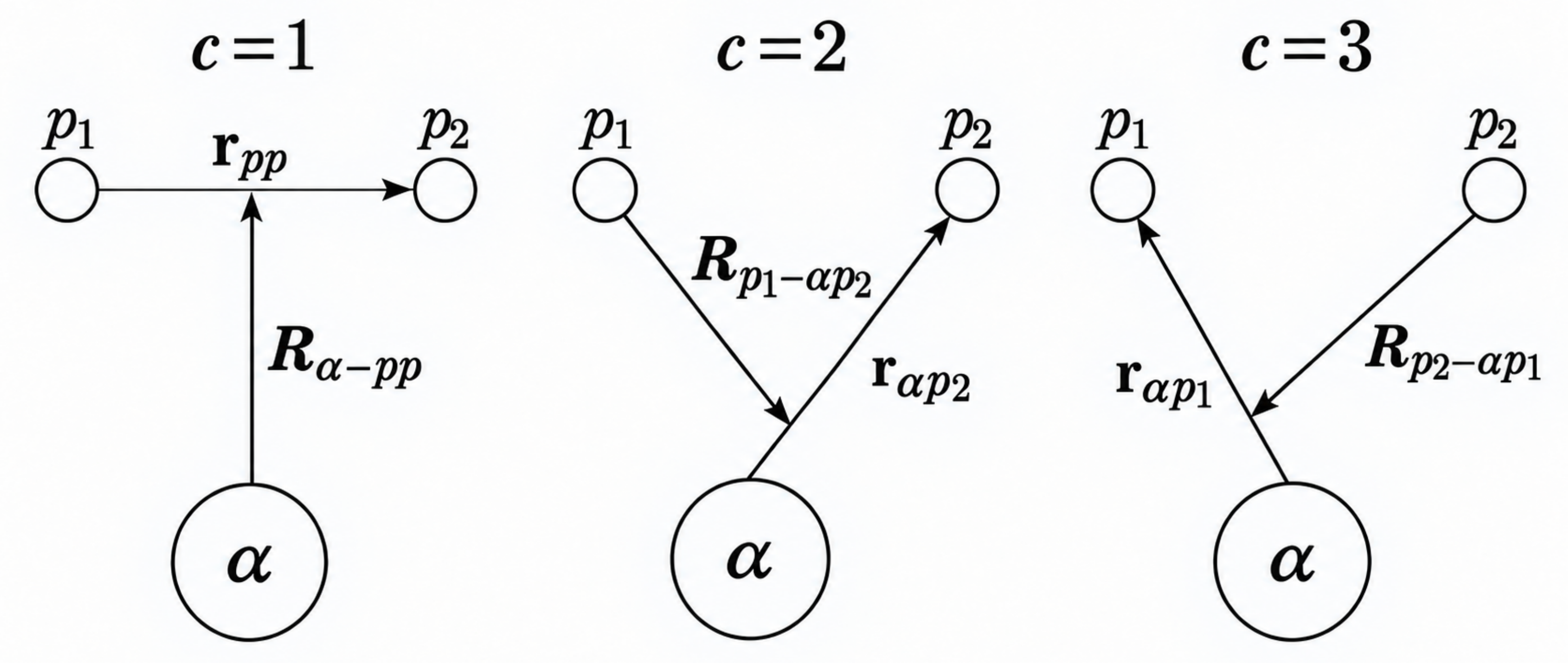}
\caption{Three Jacobi-coordinate rearrangement channels used for the $\alpha+p_1+p_2$ description of $^{6}$Be. Channel $c=1$ is the $T$-type configuration, while $c=2$ and $3$ are the two $Y$-type configurations related by exchange of the valence protons. The bold vectors denote the pair and spectator Jacobi coordinates used in the corresponding rearrangement channel.}
\label{fig:jacobi}
\end{figure}

We describe $^{6}$Be as an $\alpha+p+p$ three-body system with an inert $\alpha$ core. The Hamiltonian of the $\alpha+p+p$ three-body system is written as
\begin{align}
 \hat H_X={}&\hat T+\hat V_{\alpha p}^{(1)}+\hat V_{\alpha p}^{(2)}
 +\hat V_{pp}+\hat V_{\mathrm{Coul}}
 \nonumber\\
 &+\hat V_{\mathrm{Pauli}}+\hat V_{3b}^{(X)},
 \qquad X=\init,\finalh.
 \label{eq:hamiltonian}
\end{align}
The labels $\init$ and $\finalh$ distinguish the Hamiltonians used to prepare the initial confined state and to generate the complex-scaled eigenstates entering the ECR for the subsequent time evolution, respectively. Here, $\hat T$ is the relative kinetic energy after removing the center-of-mass motion. The proton-proton nuclear interaction is described by the realistic Argonne $v8'$ NN interaction \cite{Pudliner1997}, which contains strong short-range repulsion. The nuclear part of the $\alpha$--proton interaction is described by the KKNN potential \cite{Kanada1979}, and the Coulomb interactions are included explicitly. Pauli-forbidden components between the $\alpha$ core and the valence protons are removed by an orthogonality-condition prescription \cite{Saito1969,Kukulin1976}. In addition, we introduce a phenomenological three-body interaction with a Gaussian radial form \cite{Kikuchi2010},
\begin{equation}
 V_{3b}^{(X)}
 =V_3^{(X)}
 \exp\left[-\frac{\nu}{b_c^2}
 \left(r_{\alpha p_1}^2+r_{\alpha p_2}^2\right)\right].
 \label{eq:threebodyforce}
\end{equation}
Here, $r_{\alpha p_1}$ and $r_{\alpha p_2}$ denote the $\alpha$-proton relative distances defined by the particle labeling in Fig.~\ref{fig:jacobi}. The parameter $\nu=0.07$ is the dimensionless range parameter and $b_c=1.4$ fm is the oscillator-length parameter of the $\alpha$ core. The same radial form is used for the two Hamiltonians, and only the strength $V_3^{(X)}$ is changed.

The three-body wave functions are expanded using the Gaussian expansion method (GEM) over the three rearrangement channels $c=1,2,3$ shown in Fig.~\ref{fig:jacobi} \cite{Hiyama2003}. In the $c=1$ channel, the two valence protons form the pair, corresponding to the $T$-type Jacobi coordinates. The $c=2$ and $3$ channels are the two $Y$-type rearrangements, in which the $\alpha+p_2$ and $\alpha+p_1$ subsystems form the pair, respectively. These rearrangement channels allow both proton-proton and $\alpha$-proton correlations to be represented explicitly in the three-body wave function.

For each rearrangement channel, the orbital GEM basis is written as
\begin{align}
 \phi_{nlNL;\Lambda M}^{(c)}
 ={}&
 \mathcal N_{nlNL}^{(c)}
 \exp\left[
 -\left(\frac{r_c}{r_n^{(c)}}\right)^2
 -\left(\frac{R_c}{R_N^{(c)}}\right)^2
 \right]
 \nonumber\\
 &\times
 \left[
 \mathcal Y_l(\bm r_c)
 \mathcal Y_L(\bm R_c)
 \right]_{\Lambda M}.
 \label{eq:gem_basis}
\end{align}
Here, $\mathcal N_{nlNL}^{(c)}$ is the normalization constant of the Gaussian basis function, and $\mathcal Y_{lm}(\bm r)=r^lY_{lm}(\hat{\bm r})$ denotes a solid spherical harmonic. The Gaussian ranges $r_n^{(c)}$ and $R_N^{(c)}$ are chosen in geometric progression following the standard GEM prescription \cite{Hiyama2003}.

The total three-body wave function is constructed as a superposition of the Gaussian basis functions over all three rearrangement channels,
\begin{align}
 \Psi_{JM}
 ={}&\sum_{c=1}^{3}
 \sum_{\gamma}
 C_{\gamma}^{(c)}
 \mathcal A_{12}
 \left[
 \phi_{nlNL;\Lambda}^{(c)}(\bm r_c,\bm R_c)
 \otimes\chi_S
 \right]_{JM}.
 \label{eq:gem_wavefunction}
\end{align}
Here, $\gamma$ represents the set of indices $\{n l N L \Lambda S\}$, $(\bm r_c,\bm R_c)$ are the Jacobi relative coordinates of rearrangement channel $c$ shown in Fig.~\ref{fig:jacobi}, and $\chi_S$ is the two-proton spin wave function. The operator $\mathcal A_{12}$ antisymmetrizes the two valence protons. The summation over $c=1,2,3$ explicitly incorporates the $T$- and $Y$-type rearrangement configurations into the three-body wave function.

\subsection{Initial state and complex-scaled final states}

In the present application, the complex-scaled time evolution requires two sets of states: the initial wave function $|\Phi(0)\rangle$ and the complex-scaled eigenstates $\{|\Psi_\nu^\theta\rangle\}$ entering Eq.~(\ref{eq:timeevolution}). These states are constructed from the initial and final Hamiltonians defined in Sec.~III A, respectively.

The initial wave function $|\Phi(0)\rangle$ is prepared as a confined $0^+$ state of $^{6}$Be. 
The confined state provides a localized initial wave packet that allows the subsequent outward propagation of the decay component to be followed.
It is obtained from
\begin{equation}
 \hat H_{\init}|\Phi(0)\rangle=E_{\init}|\Phi(0)\rangle,
 \label{eq:initialstate}
\end{equation}
with an enhanced attraction of the three-body interaction so that the lowest $0^+$ state lies below the $\alpha+p+p$ threshold. In the present calculation, $V_3^{(\init)}=-7.00$ MeV gives $E_{\init}=-1.042$ MeV relative to the $\alpha+p+p$ threshold. The resulting bound $0^+$ state is used as the normalized initial wave function at $t=0$.

The complex-scaled eigenstates $|\Psi_\nu^\theta\rangle$ are obtained by diagonalizing the complex-scaled final $0^+$ Hamiltonian,
\begin{equation}
 \hat H_{\finalh}^{\theta}|\Psi_\nu^\theta\rangle
 =E_\nu^\theta|\Psi_\nu^\theta\rangle,
 \label{eq:final_csm_eigen}
\end{equation}
where the three-body strength is set to $V_3^{(\finalh)}=-0.80$ MeV so that the real part of the physical $0_1^+$ resonance energy of $^{6}$Be is reproduced. The isolated resonance eigenvalue is $E_{\mathrm{res}}^{\theta}=1.373-i\,0.053$ MeV, corresponding to a decay width of $\Gamma_{\cs}=0.106$ MeV. Only the real part of the resonance energy is used as the fitting condition for the final three-body interaction, so that the width remains a prediction of the final Hamiltonian.

The strengths of the three-body interaction and the resulting energies are summarized in Table~\ref{tab:initial_final_energy}. The confined initial state has no direct experimental counterpart and is introduced only to prepare the initial wave packet. The calculated resonance energy and width are also compared with the experimental values of Tilley \textit{et al.}~\cite{Tilley2002}.

\begin{table}[t]
\caption{Three-body interaction strengths and energies of the confined initial state and the physical $0_1^+$ resonance of $^{6}$Be. Energies are measured from the $\alpha+p+p$ threshold. The experimental resonance energy and width are also shown for comparison \cite{Tilley2002}.}
\label{tab:initial_final_energy}
\begin{ruledtabular}
\begin{tabular}{lccc}
 & $V_3^{(X)}$ (MeV) & $E$ (MeV) & $\Gamma$ (MeV) \\
\hline
Confined initial $0^+$      & $-7.00$ & $-1.042$  & --- \\
$0_1^+$ resonance (present) & $-0.80$ & $1.373$   & $0.106$ \\
$0_1^+$ resonance (exp.)    & ---     & $1.371(5)$ & $0.092(6)$ \\
\end{tabular}
\end{ruledtabular}
\end{table}

\begin{figure}[t]
\centering
\includegraphics[width=\columnwidth]{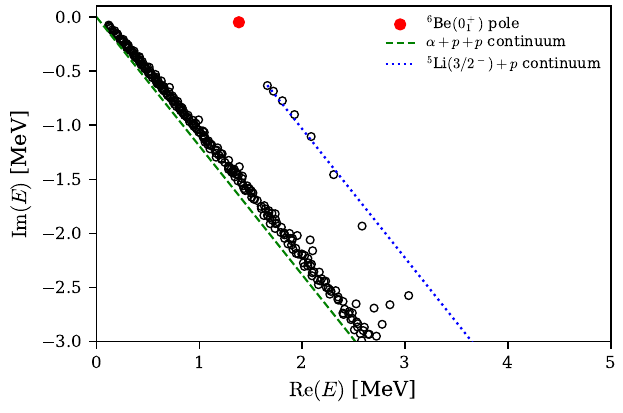}
\caption{Complex-energy spectrum $E_\nu^\theta$ of the final $0^+$ Hamiltonian for $\theta=25^\circ$. Black open circles denote the complex eigenvalues, and the red closed circle denotes the isolated $0_1^+$ resonance pole. The green dashed and blue dotted lines indicate the expected $\alpha+p+p$ and $^{5}$Li$(3/2^-)+p$ continuum branches, respectively. The latter is drawn from the $^{5}$Li$(3/2^-)$ resonance energy of $1.66-i\,0.63$ MeV calculated with the KKNN interaction.}
\label{fig:spectrum}
\end{figure}

The complex-energy spectrum is shown in Fig.~\ref{fig:spectrum} at $\theta=25^\circ$, where the $^{5}$Li$(3/2^-)+p$ continuum branch is more clearly separated from the $\alpha+p+p$ continuum. This larger scaling angle is used only to visualize the continuum structure. The time-evolution calculation is performed at $\theta=10^\circ$, where the $^{5}$Li$(3/2^-)+p$ continuum contribution is still included in the complex-scaled eigenstate expansion, although its branch is less clearly resolved in the complex-energy spectrum.

For both the confined initial state and the complex-scaled eigenstates entering the ECR, orbital angular momenta up to $l,L=2$ are included. For each Jacobi coordinate, the Gaussian ranges are taken in geometric progression over $0.25$--$15$ fm with 15 functions for the confined initial state and over $0.25$--$30$ fm with 25 functions for the complex-scaled final states.

\section{Decay observables}

\subsection{Survival probability and time-domain width}

The survival probability of the confined initial state is defined as
\begin{equation}
 P_{\surv}(t)=
 \left|\langle\Phi(0)|\Phi(t)\rangle\right|^2.
 \label{eq:survival}
\end{equation}
When the time evolution is dominated by the isolated $0_1^+$ resonance, the survival probability is expected to exhibit an approximately exponential behavior,
\begin{equation}
 P_{\surv}(t)\simeq
 A\exp\left(-\frac{\Gamma_{\td}}{\hbar}t\right).
 \label{eq:exponential}
\end{equation}
Here, $A$ accounts for the short-time nonexponential component arising from nonresonant contributions and is therefore not constrained to unity. The decay width $\Gamma_{\td}$ extracted from the exponential region is compared with the CSM pole width $\Gamma_{\cs}$.

\subsection{Projected decay component and norm loss}

In our preceding study, the norm decrease associated with the outgoing component was examined in a two-body test calculation and was found to play a role analogous to the absorption of outgoing flux by a complex absorbing potential (CAP) \cite{KikuchiTimeEvolution}. Based on this behavior, we characterize the corresponding norm decrease in the present three-body calculation. The norm of the propagated state in the physical representation is defined as
\begin{equation}
 N_{\rm phys}(t)=\langle\Phi(t)|\Phi(t)\rangle,
\end{equation}
and the norm-loss quantity is defined by
\begin{equation}
 P_{\rm loss}(t)=1-N_{\rm phys}(t).
 \label{eq:normloss}
\end{equation}
As in the preceding two-body analysis, $P_{\rm loss}(t)$ is used to characterize the outgoing component removed from the represented wave packet during the complex-scaled propagation and is not identified with the physical two-proton decay probability.

To isolate the decay component from the surviving initial-state component, we introduce the projector
\begin{equation}
 \hat P_0=|\Phi(0)\rangle\langle\Phi(0)|
\end{equation}
and define the projected decay wave function as
\begin{equation}
 |\Phi_d(t)\rangle=(1-\hat P_0)|\Phi(t)\rangle.
 \label{eq:decaywf}
\end{equation}
By construction, $\langle\Phi(0)|\Phi_d(t)\rangle=0$, and $|\Phi_d(0)\rangle=0$.

The norm of the projected decay component that remains represented in the calculation space is defined as
\begin{equation}
 P_d^{\rm rep}(t)=\langle\Phi_d(t)|\Phi_d(t)\rangle.
\end{equation}
Within the represented model space, the propagated norm is then decomposed as
\begin{equation}
 N_{\rm phys}(t)=P_{\surv}(t)+P_d^{\rm rep}(t).
 \label{eq:normbalance}
\end{equation}
Together with the definition of $P_{\rm loss}(t)$, this relation gives
\begin{equation}
 1=P_{\surv}(t)+P_d^{\rm rep}(t)+P_{\rm loss}(t).
 \label{eq:probabilitybalance}
\end{equation}
This relation separates the surviving initial component, the decay component that remains represented in the calculation space, and the norm loss during the complex-scaled propagation.

\subsection{Spatial decay density}

The coordinate-space representation of the projected decay wave function is defined as
\begin{equation}
 \Phi_d(\bm r,\bm R;t)=\langle\bm r,\bm R|\Phi_d(t)\rangle.
\end{equation}
After integrating over the angular variables, we define the two-dimensional radial decay density as
\begin{equation}
 D_d(r,R;t)=r^2R^2\int d\Omega_r\,d\Omega_R\,|\Phi_d(\bm r,\bm R;t)|^2.
 \label{eq:density}
\end{equation}
The integral of the density over the two radial coordinates gives the norm of the projected decay component,
\begin{equation}
 P_d^{\rm rep}(t)=\int_0^\infty dr\int_0^\infty dR\,D_d(r,R;t).
\end{equation}
Thus, $D_d(r,R;t)$ provides the spatial distribution of the projected decay component introduced in Sec.~IV.B.

The density is evaluated in both the $T$- and $Y$-type Jacobi-coordinate representations shown in Fig.~\ref{fig:jacobi}. The $T$-type density $D_d(r_{pp},R_{\alpha-pp};t)$ describes the distribution in the proton-proton separation $r_{pp}$ and the distance $R_{\alpha-pp}$ between the $\alpha$ core and the center of mass of the two protons. In contrast, the $Y$-type density $D_d(r_{\alpha p},R_{p-\alpha p};t)$ resolves the $\alpha+p$ subsystem through $r_{\alpha p}$, together with the motion of the spectator proton relative to this subsystem through $R_{p-\alpha p}$.

\subsection{Geometrical analysis of the decay density}

The geometry of the decay density provides useful information for characterizing different modes of three-body decay. In previous studies of $^{6}$Be, spatial distributions in Jacobi coordinates have been used to distinguish correlated two-proton motion, sequential-like configurations, and broad three-body decay patterns from their characteristic geometries \cite{Oishi2014,Oishi2017,Grigorenko2009}. Motivated by these analyses, we introduce a quantitative geometrical characterization of the projected decay density. In particular, the inner component should be separated from the outward-propagating component before the decay geometry is analyzed.

For this purpose, we introduce the mass-scaled Jacobi coordinates
\begin{equation}
 x=\sqrt{\frac{\mu_r}{m_N}}\,r,
 \qquad
 y=\sqrt{\frac{\mu_R}{m_N}}\,R,
\end{equation}
where $m_N$ is the nucleon mass and $\mu_r$ and $\mu_R$ are the reduced masses associated with the two Jacobi coordinates. The corresponding hyperradius $\rho$ and hyperangle $\beta$ are defined by
\begin{equation}
 \rho=\sqrt{x^2+y^2},
 \qquad
 \beta=\tan^{-1}\left(\frac{x}{y}\right),
 \qquad 0\leq\beta\leq\frac{\pi}{2}.
 \label{eq:hypercoords}
\end{equation}
The mass scaling provides a common geometrical measure for the $T$- and $Y$-type Jacobi coordinates while accounting for their different reduced masses.

The hyperradial distribution is obtained by integrating the projected decay density over the hyperangle,
\begin{equation}
 \mathcal P_\rho(\rho;t)
 =\int_0^{\pi/2}d\beta\,
 \frac{m_N\rho}{\sqrt{\mu_r\mu_R}}
 D_d\bigl(r(\rho,\beta),R(\rho,\beta);t\bigr),
 \label{eq:hyperradial_density}
\end{equation}
where
\begin{equation}
 r(\rho,\beta)=\sqrt{\frac{m_N}{\mu_r}}\,\rho\sin\beta,
 \qquad
 R(\rho,\beta)=\sqrt{\frac{m_N}{\mu_R}}\,\rho\cos\beta.
 \label{eq:hyperinverse}
\end{equation}
The boundary $\rho_{\rm in}$ is chosen at the minimum of $\mathcal P_\rho(\rho;t)$ that separates the inner and outer components of the projected density. Only the region $\rho>\rho_{\rm in}$ is retained in the subsequent geometrical analysis.

After removing the inner region, the remaining density is reduced to a one-dimensional hyperangular distribution,
\begin{equation}
 \mathcal P_\beta(\beta;t)
 =\int_{\rho_{\rm in}}^{\infty}d\rho\,
 \frac{m_N\rho}{\sqrt{\mu_r\mu_R}}
 D_d\bigl(r(\rho,\beta),R(\rho,\beta);t\bigr).
 \label{eq:hyperangular_density}
\end{equation}
Here, the Jacobian follows from $dr\,dR=m_N\rho\,d\rho\,d\beta/\sqrt{\mu_r\mu_R}$. With the convention in Eq.~(\ref{eq:hypercoords}), small $\beta$ corresponds to a relatively compact pair coordinate $r$ compared with the spectator coordinate $R$. The hyperangular distribution therefore characterizes how the outward-propagating decay density is partitioned between different Jacobi geometries without introducing rectangular cuts in the $(r,R)$ plane.

When two characteristic structures are present in $\mathcal P_\beta(\beta;t)$, we represent the distribution by the sum of two Gaussian components,
\begin{equation}
 \begin{aligned}
  \mathcal P_\beta(\beta;t)&\simeq\sum_{i=1}^{2}G_i(\beta;t), \\
  G_i(\beta;t)&=A_i(t)
  \exp\left[-\frac{\{\beta-\beta_i(t)\}^2}{2\sigma_i^2(t)}\right].
 \end{aligned}
 \label{eq:two_gaussian}
\end{equation}
When one component dominates the distribution, a fully unconstrained simultaneous fit can make the weaker Gaussian poorly determined. We therefore determine the Gaussian associated with the principal maximum first, using the part of the distribution on its side of the intervening minimum. This dominant component is subtracted from $\mathcal P_\beta$, and the second Gaussian is fitted to the positive residual on the opposite side. The sequential procedure reduces the ambiguity associated with a weak secondary contribution. A two-component decomposition is used quantitatively only when the secondary structure is sufficiently resolved; otherwise, its fitted parameters are not assigned a unique physical meaning.

The relative weight of each hyperangular component is defined from the integrated Gaussian area,
\begin{equation}
 f_i^{(\beta)}(t)=\frac{I_i(t)}{I_1(t)+I_2(t)},
 \qquad
 I_i(t)=\int_0^{\pi/2}d\beta\,G_i(\beta;t).
 \label{eq:hyperangular_fraction}
\end{equation}
This decomposition provides a quantitative measure of characteristic decay geometries in the outer component. The fitted Gaussian components are used as geometrical descriptors and are not identified a priori with distinct asymptotic decay channels.

\subsection{Spin decomposition and spin correlation}

The projected decay state is decomposed into the proton-proton spin-singlet and spin-triplet components as
\begin{equation}
 |\Phi_d^{S}(t)\rangle=\hat P_S|\Phi_d(t)\rangle,
 \qquad S=0,1,
\end{equation}
where $\hat P_S$ denotes the projector onto the two-proton spin $S$. The corresponding spin probabilities are defined as
\begin{equation}
 P_S(t)=\langle\Phi_d(t)|\hat P_S|\Phi_d(t)\rangle.
\end{equation}
The spin fractions normalized within the projected decay component are then given by
\begin{equation}
 f_S(t)=\frac{P_S(t)}{P_{S=0}(t)+P_{S=1}(t)}.
 \label{eq:spinfrac}
\end{equation}
The spin-resolved decay densities, $D_d^{S=0}(r,R;t)$ and $D_d^{S=1}(r,R;t)$, are evaluated in the same manner as the total decay density in Eq.~(\ref{eq:density}).

For the present $0^+$ decay, the two-proton spin correlation can be expressed directly in terms of the spin fractions as
\begin{equation}
 \langle\bm\sigma_1\cdot\bm\sigma_2\rangle_d
 =-3f_{S=0}+f_{S=1}.
 \label{eq:sigmasigma}
\end{equation}
Here, $\bm\sigma_1\cdot\bm\sigma_2$ has the eigenvalues $-3$ and $1$ for the spin-singlet and spin-triplet states, respectively.

Following the spin-correlation analysis of Oishi~\cite{Oishi2025}, the spin correlation is characterized by the Clauser-Horne-Shimony-Holt (CHSH) indicator,
\begin{equation}
 S_{\rm CHSH}(t)
 =2\sqrt{2}\left|\frac{\langle\bm\sigma_1\cdot\bm\sigma_2\rangle_d}{3}\right|.
 \label{eq:chsh_j0}
\end{equation}
A local-hidden-variable description is restricted by the Bell-CHSH bound $S_{\rm CHSH}\leq2$~\cite{CHSH1969}, whereas quantum mechanics allows the CHSH indicator to reach the Tsirelson bound $S_{\rm CHSH}=2\sqrt{2}$~\cite{Cirelson1980}. A pure spin-singlet Bell state reaches this quantum-mechanical maximum. Since $|\Phi_d(0)\rangle=0$, the normalized spin fractions and $S_{\rm CHSH}$ are undefined at $t=0$ and are evaluated only after the projected decay component becomes nonzero.

\begin{figure}[t]
\centering
\includegraphics[width=\columnwidth]{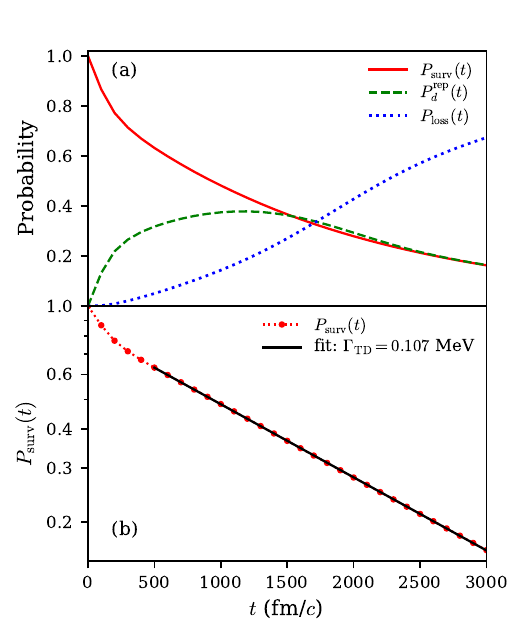}
\caption{Time evolution of the survival probability and norm-related quantities up to $3000$ fm/$c$. In panel (a), the red solid line denotes $P_{\surv}(t)$, the green dashed line denotes the represented projected-decay norm $P_d^{\rm rep}(t)$, and the blue dotted line denotes the outgoing-wave norm loss $P_{\rm loss}(t)=1-N_{\rm phys}(t)$. Panel (b) shows $P_{\surv}(t)$ on a logarithmic scale as a red dotted line with points, together with the black solid fit over $t=500$--$3000$ fm/$c$.}
\label{fig:survival}
\end{figure}

\section{Results and discussion}

\subsection{Survival probability and consistency of the decay width}

Figure~\ref{fig:survival}(a) shows the three quantities entering the norm balance of Eq.~(\ref{eq:probabilitybalance}), while panel (b) displays the survival probability on a logarithmic scale up to $3000$ fm/$c$. After the initial transient, $\ln P_{\surv}(t)$ becomes nearly linear, indicating an approximately exponential decay regime. To exclude the short-time transient, we fit the interval $t=500$--$3000$ fm/$c$ with Eq.~(\ref{eq:exponential}), obtaining $\Gamma_{\td}=0.107$ MeV. The corresponding CSM pole width is $\Gamma_{\cs}=0.106$ MeV. The close agreement between the two independently obtained widths confirms the internal consistency of the complex-scaled time-evolution framework: the late-time decay rate reproduces the resonance width determined independently from the CSM pole.

\begin{figure}[t]
\centering
\includegraphics[width=\columnwidth]{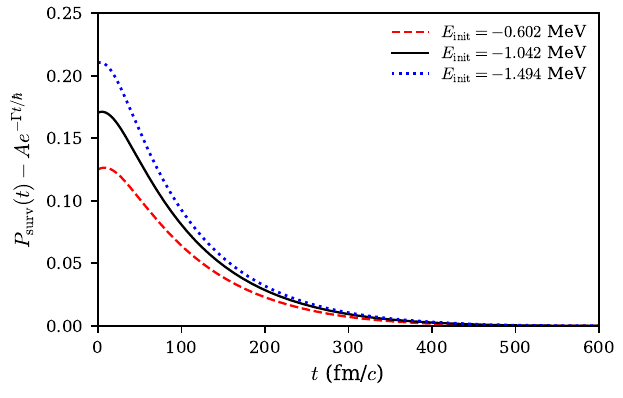}
\caption{Short-time deviation from the late-time exponential fit, $P_{\surv}(t)-Ae^{-\Gamma t/\hbar}$, for three different bound-state energies of the confined initial $0^+$ state, while the final Hamiltonian is kept fixed. The weakly bound, baseline, and more strongly bound initial states are shown by the red dashed, black solid, and blue dotted curves, respectively.}
\label{fig:shorttime_init}
\end{figure}

We next examine the short-time deviation from the late-time exponential behavior. To investigate its dependence on the preparation of the confined initial state, we performed calculations for three initial Hamiltonians producing bound $0^+$ states at $E_{\rm init}=-0.602$, $-1.042$, and $-1.494$ MeV, while keeping the final Hamiltonian unchanged. Among them, the $E_{\rm init}=-1.042$ MeV state corresponds to the baseline initial state used throughout the present calculation. Figure~\ref{fig:shorttime_init} shows the deviation
\begin{equation}
 \Delta P(t)\equiv P_{\surv}(t)-Ae^{-\Gamma t/\hbar},
\end{equation}
where $A$ and $\Gamma$ are taken from the late-time exponential fit for each initial state. The deviation is most pronounced at early times and decreases rapidly with time. Its magnitude increases as the initial state becomes more deeply bound, whereas the late-time decay width remains essentially unchanged. This contrast indicates that the short-time dynamics retain sensitivity to how the initial confined state is prepared, while the late-time decay rate is much less sensitive to it.

The dependence on the initial-state preparation is also reflected in the overlap with the $0_1^+$ resonance of the final Hamiltonian. We define
\begin{equation}
 C_{\rm res}=\langle\widetilde{\Psi}_{0_1^+}^{\theta}|\hat U(\theta)|\Phi(0)\rangle.
\end{equation}
Because the resonance wave function is complex scaled, $C_{\rm res}$ is generally complex. In a simplified resonance-dominant picture, the same resonance overlap enters the expansion of the initial state and its projection from the propagated state. The resonance contribution to the survival amplitude is therefore expected to scale approximately as $|C_{\rm res}|^2 e^{-iE_{\rm res}t/\hbar}$, giving a survival-probability contribution proportional to $|C_{\rm res}|^4 e^{-\Gamma t/\hbar}$. Accordingly, the fitted prefactor $A$ is expected to approximately follow $|C_{\rm res}|^4$.

As shown in Table~\ref{tab:initial_overlap}, the fitted values of $A$ closely follow $|C_{\rm res}|^4$ for all three initial states. As the initial state becomes more deeply bound, both quantities decrease, while the short-time deviation in Fig.~\ref{fig:shorttime_init} becomes larger. The remaining difference between $A$ and $|C_{\rm res}|^4$ reflects contributions beyond an isolated-resonance description.

The observed short-time sensitivity is consistent with the analysis of Wang \textit{et al.}~\cite{Wang2023}, who emphasized that nonexponential decay can retain information on the structure and preparation of the initial state. Sensitivity of $^{6}$Be three-body correlations to the initial structure and reaction mechanism has also been discussed by Grigorenko \textit{et al.}~\cite{Grigorenko2012}. In the present calculation, changing only the confinement of the initial state modifies both the early-time deviation and the resonance content, while leaving the late-time decay width nearly unchanged. Thus, the present three-body calculation is consistent with the general picture obtained in previous studies: the early-time behavior is sensitive to how the initial confined state is prepared, whereas the late-time decay rate remains nearly unchanged.

\begin{table}[h!]
\caption{Resonance overlap amplitudes $C_{\rm res}$, the corresponding $|C_{\rm res}|^4$, and the fitted exponential prefactors $A$ for the three confined initial $0^+$ states.}
\label{tab:initial_overlap}
\begin{ruledtabular}
\begin{tabular}{cccc}
$E_{\rm init}$ (MeV) & $C_{\rm res}$ & $|C_{\rm res}|^4$ & $A$ \\
\hline
$-0.602$ & $0.961-i\,0.051$ & 0.858 & 0.875 \\
$-1.042$ & $0.948-i\,0.055$ & 0.813 & 0.829 \\
$-1.494$ & $0.937-i\,0.058$ & 0.777 & 0.790 \\
\end{tabular}
\end{ruledtabular}
\end{table}

\begin{figure}[!htb]
\centering
\includegraphics[width=\columnwidth]{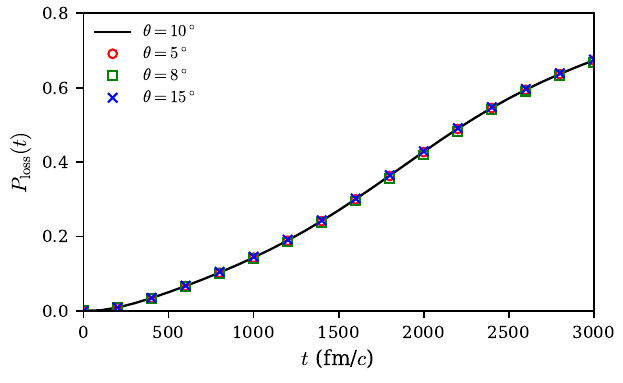}
\caption{Norm loss $P_{\rm loss}(t)=1-N_{\rm phys}(t)$ for four complex-scaling angles. The result for $\theta=10^\circ$, used as the reference scaling angle in the present calculation, is shown by the black solid line; those for $\theta=5^\circ$, $8^\circ$, and $15^\circ$ are shown by red open circles, green open squares, and blue crosses, respectively.}
\label{fig:ploss_theta}
\end{figure}

\begin{figure*}[t]
\centering
\includegraphics[width=0.98\textwidth]{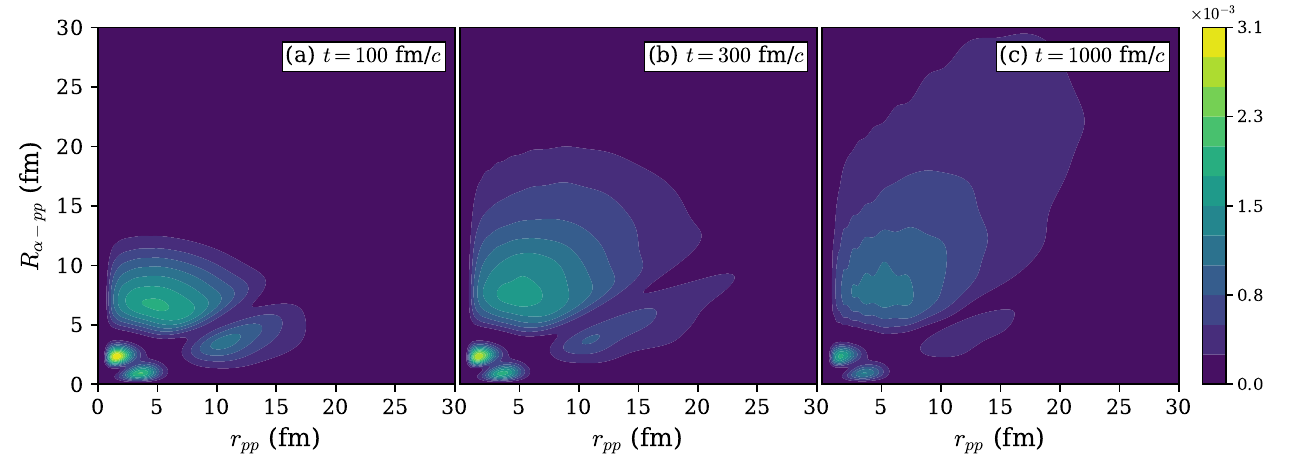}
\caption{Time evolution of the projected decay density $D_d(r_{pp},R_{\alpha-pp};t)$ in the $T$-type Jacobi coordinates. Panels (a), (b), and (c) correspond to $t=100$, 300, and 1000 fm/$c$, respectively. The initial-state component is removed according to Eq.~(\ref{eq:decaywf}). A common linear color scale is used for all three panels.}
\label{fig:Tdensity}
\end{figure*}

\begin{figure*}[t]
\centering
\includegraphics[width=0.98\textwidth]{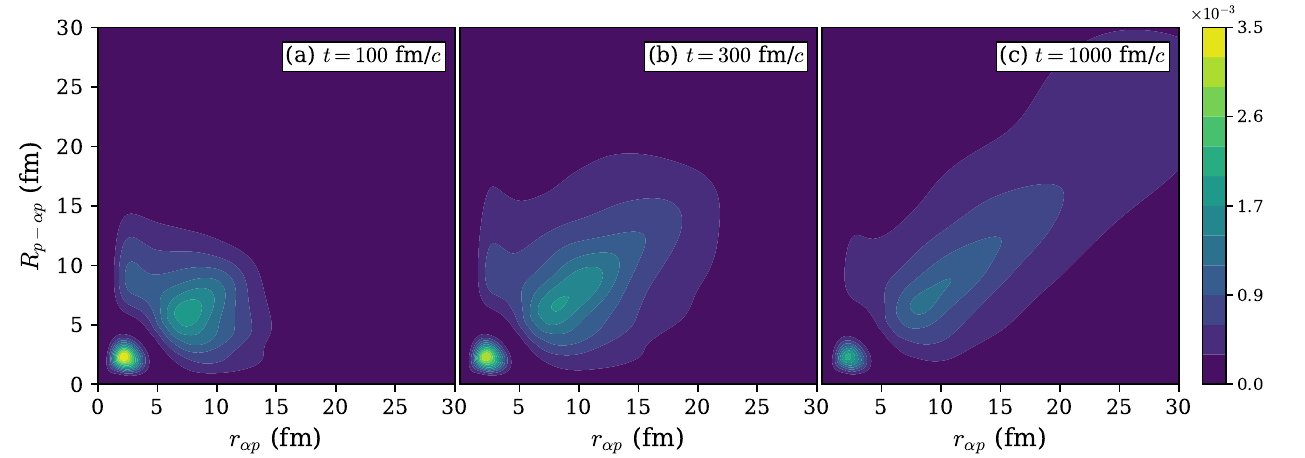}
\caption{Same as Fig.~\ref{fig:Tdensity}, but for the $Y$-type Jacobi coordinates $D_d(r_{\alpha p},R_{p-\alpha p};t)$. Panels (a), (b), and (c) correspond to $t=100$, 300, and 1000 fm/$c$, respectively.}
\label{fig:Ydensity}
\end{figure*}

\subsection{Norm loss and outgoing-wave propagation}

We next examine the norm loss associated with outgoing-wave propagation in the complex-scaled representation. As defined in Eq.~(\ref{eq:normloss}), $P_{\rm loss}(t)=1-N_{\rm phys}(t)$ measures the decrease of the represented norm in the physical representation and is distinct from the survival probability. As discussed in our preceding two-body study~\cite{KikuchiTimeEvolution}, this norm decrease plays a role analogous to the absorption of outgoing flux by a complex absorbing potential. In the present three-body calculation, $P_{\rm loss}(t)$ therefore provides a measure of the outgoing component removed from the represented wave packet during the complex-scaled propagation and should not be interpreted as the physical two-proton decay probability.

Figure~\ref{fig:ploss_theta} shows $P_{\rm loss}(t)$ for $\theta=5^\circ$, $8^\circ$, $10^\circ$, and $15^\circ$. The four curves remain close over the full propagation interval, demonstrating that the removal of the outgoing component depends only weakly on the complex-scaling angle over the range considered. This weak angle dependence is consistent with the CAP-like interpretation found in the preceding two-body calculation and indicates that the outgoing component is continuously removed as the wave packet propagates to large distances. This continuous loss of outgoing strength is important for interpreting the late-time spatial densities discussed below.

\subsection{Time-dependent spatial decay pattern}

We next examine the spatial evolution of the projected decay component using the density defined in Eq.~(\ref{eq:density}). Figure~\ref{fig:Tdensity} shows the $T$-type projected decay density $D_d(r_{pp},R_{\alpha-pp};t)$ at $t=100$, 300, and 1000 fm/$c$. At $t=100$ fm/$c$, appreciable strength is already concentrated at relatively small proton-proton separations while extending toward larger $R_{\alpha-pp}$. As time proceeds, the density propagates outward, but a substantial component remains localized at comparatively small $r_{pp}$. The persistence of the small-$r_{pp}$ component is consistent with previous time-dependent studies that emphasized correlated two-proton emission in $^{6}$Be. Importantly, the same qualitative feature survives in the present calculation despite the strong short-range repulsion of the realistic Argonne $v8'$ NN interaction and the explicit treatment of the Jacobi rearrangement channels.

The $Y$-type projected density $D_d(r_{\alpha p},R_{p-\alpha p};t)$ is shown in Fig.~\ref{fig:Ydensity}. At early times, appreciable strength appears at relatively small $r_{\alpha p}$, indicating a component with a compact $\alpha+p$ geometry. As time proceeds, the density develops over a much broader region of both $r_{\alpha p}$ and $R_{p-\alpha p}$. Together with the $T$-type result, this evolution shows how the two rearranged Jacobi representations provide complementary information on the decay geometry: the correlated two-proton component remains prominent, whereas the compact $\alpha+p$ configuration becomes progressively less important.

At later times, the represented spatial density approaches a quasi-stationary pattern within the analysis region. This apparent saturation does not imply that the physical outgoing wave has stopped propagating. Rather, components reaching larger distances are continuously removed from the represented wave packet by the complex-scaled propagation, as reflected in the norm loss discussed in Sec.~V.B.

\subsection{Quantitative geometrical decomposition}

\begin{figure}[t]
\centering
\includegraphics[width=\columnwidth]{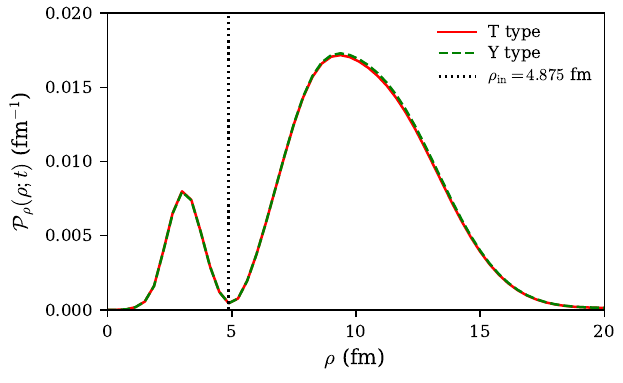}
\caption{Mass-scaled hyperradial distributions of the projected decay density at $t=100$ fm/$c$ in the $T$- and $Y$-type Jacobi coordinates. The $T$- and $Y$-type results are shown by the red solid and green dashed curves, respectively. The black dotted vertical line indicates the common minimum $\rho_{\rm in}=4.875$ fm, which separates the inner and outer components of the projected decay density used in the subsequent hyperangular analysis.}
\label{fig:hyperradial}
\end{figure}

We next quantify the geometrical components identified in the two-dimensional densities using the hyperradial and hyperangular distributions introduced in Sec.~IV.D. Figure~\ref{fig:hyperradial} shows the mass-scaled hyperradial distributions at $t=100$ fm/$c$ for the $T$- and $Y$-type Jacobi coordinates. The two distributions are nearly identical and exhibit a common minimum at $\rho=4.875$ fm separating the inner and outer components. We therefore adopt $\rho_{\rm in}=4.875$ fm for both Jacobi representations and analyze the outer component with $\rho>\rho_{\rm in}$.

\begin{figure}[t]
\centering
\includegraphics[width=\columnwidth]{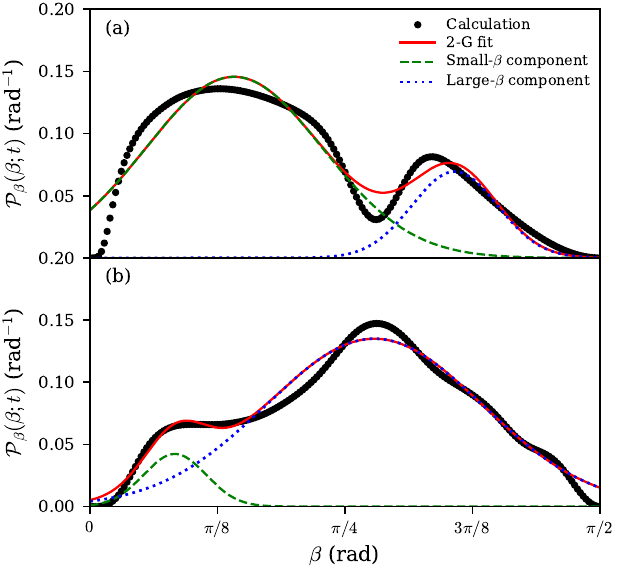}
\caption{Hyperangular distributions of the projected decay density at $t=100$ fm/$c$ after excluding $\rho\leq\rho_{\rm in}$. Panels (a) and (b) show the $T$- and $Y$-type Jacobi coordinates, respectively. Black closed circles denote the calculation, the red solid curves show the two-Gaussian fit, the green dashed curves show the small-$\beta$ component, and the blue dotted curves show the large-$\beta$ component. Small $\beta$ corresponds to a relatively compact pair coordinate, namely the $pp$ pair in the $T$-type representation and the $\alpha+p$ pair in the $Y$-type representation.}
\label{fig:hyperangular}
\end{figure}

Figure~\ref{fig:hyperangular} shows the corresponding hyperangular distributions at $t=100$ fm/$c$. In the $T$-type representation, the small-$\beta$ component carries a fraction of $0.804$, whereas the large-$\beta$ component contributes $0.196$. The dominance of the small-$\beta$ component quantitatively confirms that the outward-propagating density is largely associated with a relatively compact proton-proton configuration.

In the $Y$-type representation, the large-$\beta$ component dominates with a fraction of $0.916$, while the small-$\beta$ component contributes only $0.084$. The latter corresponds to a relatively compact $\alpha+p$ geometry with a more distant spectator proton and may therefore be regarded as a sequential-like geometry. This geometrical component is not identified exclusively with the $^{5}$Li$(3/2^-)+p$ continuum channel. The much larger large-$\beta$ fraction shows that such a compact $\alpha+p$ geometry constitutes only a minor component of the represented outer density. These fitted fractions characterize geometrical components of the represented outer density and should not be interpreted as asymptotic branching fractions.

To follow the geometrical evolution quantitatively, the same decomposition is applied at $t=100$, 300, and 1000 fm/$c$, corresponding to the snapshots shown in Figs.~\ref{fig:Tdensity} and \ref{fig:Ydensity}. The resulting fractions are summarized in Table~\ref{tab:geometry_fraction}. In the $T$-type coordinates, the small-$\beta$ fraction remains within $0.80$--$0.84$, demonstrating that the correlated two-proton configuration persists during the outward propagation. In contrast, the $Y$-type small-$\beta$ fraction decreases from $0.084$ at $t=100$ fm/$c$ to $0.038$ at $t=1000$ fm/$c$, while the large-$\beta$ component becomes correspondingly more dominant. Thus, the relatively compact $\alpha+p$ configuration becomes progressively less important as the broad three-body configuration develops. This contrasting behavior provides a quantitative counterpart to the complementary decay patterns seen directly in the two rearranged Jacobi-coordinate densities.

\begin{table}[h!]
\caption{Fractions of the small- and large-$\beta$ geometrical components obtained from the two-Gaussian decomposition at the representative times used in Figs.~\ref{fig:Tdensity} and \ref{fig:Ydensity}. The cutoff $\rho_{\rm in}=4.875$ fm is kept fixed at all three times. These fractions characterize geometrical components of the represented outer density and are not asymptotic branching fractions.}
\label{tab:geometry_fraction}
\begin{ruledtabular}
\begin{tabular}{ccccc}
$t$ & \multicolumn{2}{c}{$T$ type} & \multicolumn{2}{c}{$Y$ type} \\
(fm/$c$) & small-$\beta$ & large-$\beta$ & small-$\beta$ & large-$\beta$ \\
\hline
100  & 0.804 & 0.196 & 0.084 & 0.916 \\
300  & 0.838 & 0.162 & 0.076 & 0.924 \\
1000 & 0.818 & 0.182 & 0.038 & 0.962 \\
\end{tabular}
\end{ruledtabular}
\end{table}

\subsection{Spin-singlet correlation and CHSH indicator}

\begin{figure}[t]
\centering
\includegraphics[width=0.78\columnwidth]{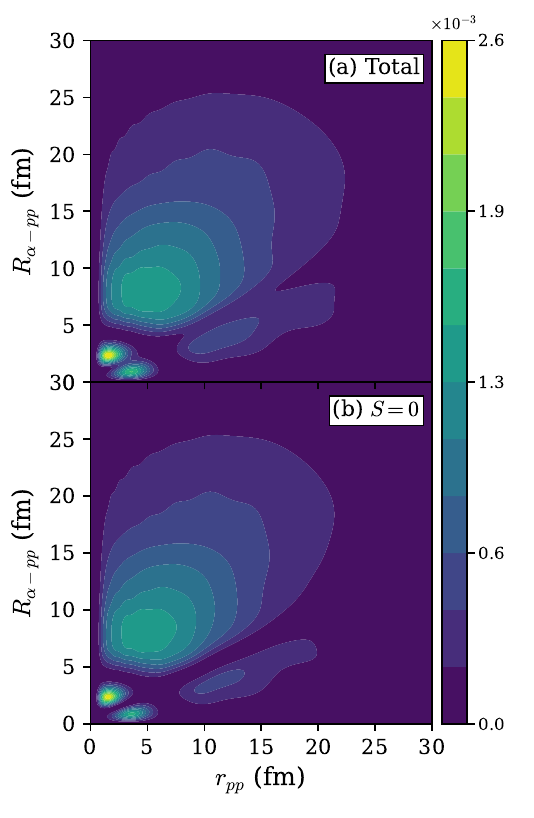}
\caption{Representative $T$-type projected decay densities at $t=500$ fm/$c$. Panels (a) and (b) show the total projected density $D_d(r_{pp},R_{\alpha-pp};t)$ and its $S=0$ component, respectively. The two panels are shown with the same linear color scale to facilitate direct comparison.}
\label{fig:spin_density}
\end{figure}

\begin{figure}[t]
\centering
\includegraphics[width=0.92\columnwidth]{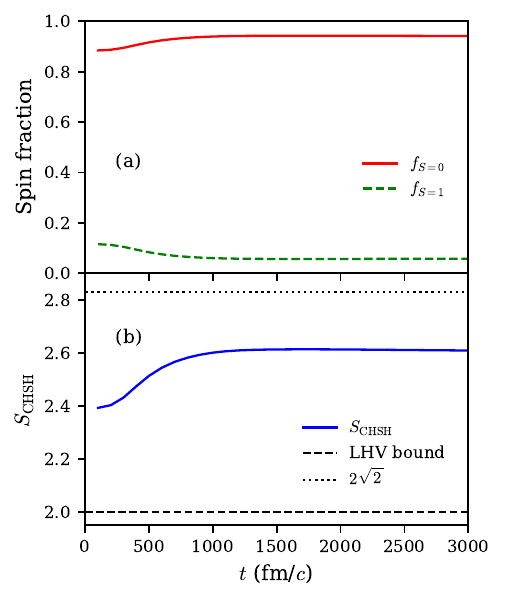}
\caption{Long-time evolution of the spin observables up to $3000$ fm/$c$. In panel (a), the red solid line denotes the spin-singlet fraction $f_{S=0}$ and the green dashed line denotes the spin-triplet fraction $f_{S=1}$ of the projected decay component. In panel (b), the blue solid line denotes the CHSH indicator defined in Eq.~(\ref{eq:chsh_j0}); the black dashed line denotes the local-hidden-variable bound $S_{\rm CHSH}=2$, and the black dotted line denotes the Tsirelson bound $S_{\rm CHSH}=2\sqrt{2}$.}
\label{fig:spin}
\end{figure}

Figure~\ref{fig:spin} shows the time evolution of the spin fractions and the corresponding CHSH indicator up to $3000$ fm/$c$. The projected decay component is strongly dominated by the proton-proton spin-singlet sector already at early times, and the singlet fraction increases further as the decay proceeds. Correspondingly, the CHSH indicator exceeds the local-hidden-variable bound $S_{\rm CHSH}=2$ and approaches a nearly constant value of about 2.61 at late times. The late-time plateau reflects the stabilization of the normalized spin composition of the represented decay component. This behavior is consistent with the time-dependent spin-correlation analysis of Ref.~\cite{Oishi2025}, despite the different proton-proton interaction and spatial representation used here. The agreement indicates that the dominant spin-singlet correlation is not strongly altered by introducing explicit Jacobi rearrangement and the realistic Argonne $v8'$ NN interaction.

The spin-resolved densities provide a direct spatial counterpart to this spin-singlet dominance. The $S=0$ component is concentrated at smaller proton-proton separations than the much weaker $S=1$ contribution. As shown in Fig.~\ref{fig:spin_density}, the total projected density and the $S=0$ density at $t=500$ fm/$c$ have very similar spatial structures. This correspondence shows that the small-$r_{pp}$ proton-proton configuration identified in the $T$-type coordinates is predominantly associated with the spin-singlet component.

The spatial and spin analyses consistently indicate that the emitted wave packet retains a dominant correlated two-proton component during the decay, while the relatively compact $\alpha+p$ configuration becomes less important with time. The persistence of the correlated two-proton component and the dominant spin-singlet character are consistent with the qualitative picture obtained in previous time-dependent studies \cite{Oishi2014,Oishi2017,Oishi2025}. The present calculation extends that picture in two respects: the explicit comparison of different Jacobi rearrangements provides complementary information on the evolving three-body geometry, and the correlated two-proton structure remains visible even with the strong short-range repulsion of the realistic Argonne $v8'$ NN interaction.

\section{Summary}

We have applied our complex-scaled time-evolution operator to the two-proton decay of $^{6}$Be in an $\alpha+p+p$ three-body model. The confined initial $0^+$ state is expanded over the complex-scaled eigenstates of the final Hamiltonian and propagated through the extended completeness relation. The survival probability becomes approximately exponential after the short-time transient, and the decay width extracted from the late-time evolution, $\Gamma_{\td}=0.107$ MeV, agrees closely with the CSM pole width, $\Gamma_{\cs}=0.106$ MeV. In contrast, the short-time nonexponential behavior is sensitive to how the initial confined state is prepared, while the late-time decay rate remains nearly unchanged. This behavior is consistent with previous studies emphasizing the sensitivity of nonexponential decay to the initial-state structure \cite{Wang2023}. The norm loss reflects the continuous removal of outgoing strength from the represented wave packet during the complex-scaled propagation and is distinct from the physical two-proton decay probability.

The projected decay densities in the $T$- and $Y$-type Jacobi coordinates reveal complementary aspects of the three-body decay geometry. In the $T$-type representation, the outward-propagating density retains a substantial component at relatively small proton-proton separation. The hyperangular decomposition quantifies this behavior: the small-$\beta$ component carries about $80\%$--$84\%$ of the represented outer density at $t=100$, 300, and 1000 fm/$c$. Thus, a correlated two-proton configuration persists throughout the propagation. In the $Y$-type representation, the small-$\beta$ component associated with a relatively compact $\alpha+p$ geometry decreases from about $8\%$ at $100$ fm/$c$ to about $4\%$ at $1000$ fm/$c$, while the large-$\beta$ component becomes correspondingly dominant. The explicit comparison of the two Jacobi rearrangements therefore provides complementary information that is not available from a single coordinate representation. These fractions characterize geometrical components of the represented outer density and should not be interpreted as asymptotic branching fractions.

The projected decay component is also strongly dominated by the proton-proton spin-singlet sector. The CHSH indicator remains above the local-hidden-variable bound and approaches a late-time value of about $2.61$, while the spin-resolved densities show that the small-$r_{pp}$ proton-proton structure is predominantly associated with the $S=0$ component. The persistence of the dominant spin-singlet component is consistent with previous time-dependent studies of $^{6}$Be~\cite{Oishi2025}. Together with the spatial analysis, the present results show that the characteristic correlated two-proton picture persists in a three-body description with explicit Jacobi-coordinate rearrangement and the realistic Argonne $v8'$ NN interaction, despite its strong short-range repulsion. In this sense, the present calculation is consistent with the essential conclusions of earlier time-dependent studies while extending them to a different and more explicit three-body framework. The present study also demonstrates that our complex-scaled time-evolution operator can be applied successfully to an explicit three-body decay problem and provides a common framework for describing the decay width, spatial geometry, and spin correlations in $^{6}$Be two-proton decay.

\begin{acknowledgments}
The work shown in this paper was supported by JSPS KAKENHI Grants No. JP26K07092, No. JP25H01268, and JST ERATO Grant No. JPMJER2304, Japan.
\end{acknowledgments}

\end{document}